\documentclass[conference,a4paper]{IEEEtran}

\usepackage[nolist]{acronym}
\usepackage{amssymb}
\usepackage{bm}
\usepackage{mathrsfs}
\usepackage[cmex10]{amsmath}  % Use the [cmex10] option to ensure complicance
\usepackage{xcolor}
\usepackage[ruled,vlined,titlenumbered,linesnumbered]{algorithm2e}
\usepackage{setspace}
\usepackage{theorem}

\usepackage{tikz} % To generate the plot from csv
\usepackage{pgfplots}
\pgfplotsset{compat=newest} % Allows to place the legend below plot
\usepackage{mathtools}
\mathtoolsset{showonlyrefs}

\newtheorem{theorem}{Theorem}
\newtheorem{definition}{Definition}
\newtheorem{lemma}{Lemma}

\newtheorem{problem}{Problem}

\newcommand{\mycode}[1]{\ensuremath{\mathcal{#1}}}
\newcommand{\myspace}[1]{\mathcal{#1}}
\newcommand{\Rowspace}[1]{\ensuremath{\myspace{R}_{q}\!\left(#1\right)}}

\newcommand{\Fq}{\ensuremath{\mathbb F_{q}}}
\newcommand{\Fqm}{\ensuremath{\mathbb F_{q^m}}}

\renewcommand{\vec}[1]{\ensuremath{\bm{#1}}}
\newcommand{\mat}[1]{\ensuremath{\bm{#1}}}

\DeclareMathOperator{\wt}{wt}

\DeclareMathOperator{\rkq}{rk_{q}}

\DeclareMathOperator{\DEC}{DEC}

\DeclareMathOperator{\NSR}{NSR}

\DeclareMathOperator{\uDR}{\tau}

\DeclareMathOperator{\lbWgen}{\tilde{W}_{gen}^{(LB)}}
\DeclareMathOperator{\ubWgen}{\tilde{W}_{gen}^{(UB)}}

\DeclareMathOperator{\lbWrand}{\tilde{W}_{RD}^{(LB)}}
\DeclareMathOperator{\ubWrand}{\tilde{W}_{RD}^{(UB)}}

\DeclareMathOperator{\WrandLB}{W_{RD}^{(LB)}}
\DeclareMathOperator{\WrandUB}{W_{RD}^{(UB)}}
\DeclareMathOperator{\Wropt}{W_{opt}^{(UB)}}
\DeclareMathOperator{\Wsim}{W_{sim}}
\DeclareMathOperator{\wcomp}{\psi}

\DeclareMathOperator{\ext}{ext}
\DeclareMathOperator{\diag}{diag}
\DeclareMathOperator*{\argmax}{arg\,max}

\DeclareMathOperator{\NM}{NM}

\DeclareMathOperator{\prob}{\varphi}

\newcommand{\shots}{\ensuremath{\ell}}

\newcommand{\RS}{\ensuremath{\mathcal{E}_{R}}}
\newcommand{\CS}{\ensuremath{\mathcal{E}_{C}}}
\newcommand{\RSv}{\ensuremath{\bm{\mathcal{E}}_{R}}}
\newcommand{\CSv}{\ensuremath{\bm{\mathcal{E}}_{C}}}
\newcommand{\ES}{\ensuremath{{\mathcal{E}}}}
\newcommand{\GS}{\ensuremath{{\mathcal{U}}}}
\newcommand{\ESv}{\ensuremath{\bm{\mathcal{E}}}}
\newcommand{\GSv}{\ensuremath{\bm{\mathcal{U}}}}

\newcommand{\SumRankWeight}{\ensuremath{\wt_{\Sigma R}^{(\n)}}}

\newcommand{\SumRankDist}{d_{\ensuremath{\Sigma}R}^{(\n)}}

\DeclareMathOperator{\LRS}{LRS}

\newcommand{\LRSCode}{\ensuremath{\mathcal{C}_{\LRS}}}

\newcommand{\errWeight}{\ensuremath{w}}
\newcommand{\decRadius}{\ensuremath{t}}
\newcommand{\gsWeight}{\ensuremath{u}}
\newcommand{\errWeightVec}{\ensuremath{\w}}
\newcommand{\gsWeightVec}{\ensuremath{\u}}

\newcommand{\sdInter}{\ensuremath{\epsilon}}

\newcommand{\NSRM}[4]{\NSR_{#1}(#2, #3, #4)}

\newcommand{\todo}[1]{}
\newcommand{\tj}[1]{}
\newcommand{\hb}[1]{}
\newcommand{\fh}[1]{}
\newcommand{\aw}[1]{}

\newcommand{\revision}[1]{#1}

\newcommand{\sample}{\overset{\$}{\gets}}
\newcommand{\defeq}{:=}

\newcommand{\Conv}{%
  \mathop{\scalebox{1.5}{\raisebox{-0.2ex}{$\circledast$}}
  }
}

\newcommand{\conv}{\ensuremath{\circledast}}

\newcommand{\Grassm}[3]{\myspace{G}_{#1}(#2,#3)}

\DeclareMathOperator{\sumDim}{\ensuremath{\dim_{\Sigma}}}

\newcommand{\quadbinom}[2]{\left[\genfrac{}{}{0pt}{}{#1\vphantom{N_N}}{#2\vphantom{N}}\right]_{q}}

\newcommand{\wdecomp}[3]{\ensuremath{\mathcal{T}_{#1,#2,#3}}}

\renewcommand{\a}{\vec{a}}
\renewcommand{\b}{\vec{b}}
\renewcommand{\c}{\vec{c}}
\newcommand{\e}{\vec{e}}

\newcommand{\n}{\vec{n}}

\newcommand{\s}{\vec{s}}

\renewcommand{\u}{\vec{u}}

\newcommand{\w}{\vec{w}}
\newcommand{\x}{\vec{x}}
\newcommand{\y}{\vec{y}}

\newcommand{\A}{\mat{A}}
\newcommand{\B}{\mat{B}}

\usepackage{algpseudocode}

\algnewcommand\AND{\textbf{and}}

\newcommand{\oh}[1]{\bnd{O}{#1}}

\newcommand{\bnd}[2]{\ensuremath{#1\mathopen{}\left(#2\right)\mathclose{}}}

\begin{acronym}
 \acro{BCH}{Bose--Chaudhuri--Hocquenghem}
 \acro{BMD}{bounded minimum distance}
 \acro{ESP}{error span polynomial}
 \acro{ELP}{error locator polynomial} 
 \acro{SRS}{skew Reed--Solomon}
 \acro{ISRS}{interleaved skew Reed--Solomon}
 \acro{lclm}{least common left multiple}
 \acro{LRS}{linearized Reed--Solomon}
 \acro{LLRS}{lifted linearized Reed--Solomon}
 \acro{ILRS}{interleaved linearized Reed--Solomon}
 \acro{LILRS}{lifted interleaved linearized Reed--Solomon}
 \acro{MDS}{maximum distance separable}
 \acro{MSRD}{maximum sum-rank distance}
 \acro{MSD}{maximum skew distance}
 \acro{RS}{Reed--Solomon}
 \acro{REF}{row echelon form}
 \acro{LEEA}{linearized extended Euclidean algorithm}
 \acro{SEEA}{skew extended Euclidean algorithm}
 \acro{VSD}{vector-symbol decoding}
 \acro{AG}{algebraic-geometry}
 \acro{KEM}{key encapsulation mechanism}
 \acro{ISD}{information-set-decoding}
 \acro{SRMD}{sum-rank-metric decoding}
 \acro{pmf}{probability mass function}
 \acro{p.m.f.}{probability mass function}
 \acro{PMF}{probability mass function}
 \acro{ML}{maximum-likelihood}
\end{acronym}

\usepackage[utf8]{inputenc} 
\usepackage[T1]{fontenc}
\usepackage{url}              % provides \url{...}
\usepackage{cite}             % improves presentation of citations

\usepackage{mleftright}       % fix to wrong spacing of \left-,
\mleftright                   % \middle- \right-commands 

\usepackage{graphicx}         % provides \includegraphics{...} to
\usepackage{booktabs}         % fixes poor spacing in tables and
 \IEEEoverridecommandlockouts
\begin{document}

\title{Randomized Decoding of Linearized Reed--Solomon Codes Beyond the Unique Decoding Radius} 

%%%%%%
% \author{%
%   \IEEEauthorblockN{Anonymous Authors}
%   \IEEEauthorblockA{%
%     Please do NOT provide authors' names and affiliations\\
%     in the paper submitted for review, but keep this placeholder.\\
%     ISIT23 follows a \textbf{double-blind reviewing policy}.}
% \thanks{This work was supported by the German Research Council (DFG) as an ANR-DFG project under Grant no. WA 3907/9-1}
% }

\author{
\IEEEauthorblockN{Thomas Jerkovits, Hannes Bartz}
\IEEEauthorblockA{\textit{Institute of Communication and Navigation} \\
\textit{German Aerospace Center (DLR)}\\
% Oberpfaffenhofen-Wessling, Germany \\
\{thomas.jerkovits, hannes.bartz\}@dlr.de}
\and
\IEEEauthorblockN{Antonia Wachter-Zeh}
\IEEEauthorblockA{\textit{School of Computation, Information and Technology} \\
\textit{Technical University of Munich (TUM)}\\
% Munich, Germany \\
antonia.wachter-zeh@tum.de}
 \thanks{This work was supported by the German Research Council (DFG) as an ANR-DFG project under Grant no. WA 3907/9-1.}
}

\maketitle

%%%%%
%% Abstract: 
%% If your paper is eligible for the student paper award, please add
%% the comment "THIS PAPER IS ELIGIBLE FOR THE STUDENT PAPER
%% AWARD." as a first line in the abstract. 
%% For the final version of the accepted paper, please do not forget
%% to remove this comment!
%%
\begin{abstract}
% THIS PAPER IS ELIGIBLE FOR THE STUDENT PAPER AWARD.
%  Instructions are given for the preparation and submission of papers
%  for the \emph{2023 International Symposium on Information
%    Theory}. This template is available (including \LaTeX{}-source)
%  from \url{https://isit2023.org/}

In this paper we address the problem of decoding \ac{LRS} codes beyond their unique decoding radius. 
We analyze the complexity in order to evaluate if the considered problem is of cryptographic relevance, i.e., can be used to design cryptosystems that are computationally hard to break.
We show that our proposed algorithm improves over other generic algorithms that do not take into account the underlying code structure.  
\end{abstract}

\acresetall
\section{Introduction}
\label{sec:introduction}
The sum-rank metric is a generalization of both, the Hamming and the rank metric, and was first introduced for space-time codes in~\cite{lu2005unified}.
Since then, several code constructions and decoders have been proposed for the sum-rank metric~\cite{Wachter2011punit,wachter2012rmcrandomlinear,wachter2015convrank,napp2017MRD,martinez2018skew,martinez2019reliable,bartz2021fastdec,bartz2022fast}.
\Ac{LRS} codes were later introduced by Mart\'{\i}nez-Pe\~nas which include Reed--Solomon codes and Gabidulin codes as special cases~\cite{martinez2019univ}.
% \Ac{LRS} codes fullfil the Singleton-like bound in the sum-rank metric with equality.
\Ac{LRS} codes are of interest for applications such as multishot network coding~\cite{nobrega2010multishot,martinez2019reliable}, locally repairable codes~\cite{martinez2019univ}, space-time codes~\cite{lu2005unified}, and code-based quantum-resistant cryptography~\cite{puchinger2022generic}.

It is well-known that the problem of decoding beyond the unique decoding radius, specifically \emph{maximum-likelihood} decoding, is a difficult problem w.r.t. the complexity.
For the Hamming metric, many works have investigated the hardness of this problem~\cite{Berlekamp1978codingproblems,stern1993npcomplete,vardy1997intract}.
List decoding is another method to decode beyond the unique decoding radius and the complexity depends on the list size. Bounds on the list size for \ac{LRS} codes are given in~\cite{Puchinger2021listLRS} and it was shown that some families of \ac{LRS} codes cannot be decoded beyond the unique decoding radius.
\revision{The exponential complexity of list decoding makes it a potentially useful tool for cryptography. Before designing cryptosystems based on the list decoding problem in the sum-rank metric, its computational complexity must be carefully studied and analyzed.}
For the rank metric the problem of decoding beyond the unique decoding radius was addressed in~\cite{renner2020randomized} for Gabidulin codes.
\revision{
Known structural attacks for McEliece-like cryptosystem in the Hamming and Rank metric~\cite{overbeck2008structural,wieschebrink2006attack,sidelnikov1992insecurity} have been generalized to the sum-rank metric~\cite{cbchoermann2022}.
This raises the question if the sum-rank metric can be adapted to other cryptosystems that are based on the hardness of decoding beyond the unique decoding radius, such as~\cite{faure2006new, wachterzeh2018repairing,LIGA-Journal}.}

In this paper we propose an algorithm which generalizes the one from~\cite{renner2020randomized} to \ac{LRS} codes (in the sum-rank metric).
\revision{Note that the work factor, i.e., the computational complexity of the algorithm, derived in~\cite{renner2020randomized} can be used to assess the security level of cryptosystems like~\cite{faure2006new, wachterzeh2018repairing,LIGA-Journal}. Thus, the work factors derived in this paper might be used to assess the security level of similar cryptosystems in the sum-rank metric.}
%\hb{I suggest to mention the points above in a more structured way: first we say that list decoding of LRS codes is hard, then we jump to the generalization of an algorithm in the rank metric. Maybe add some explicit decoders first, then say that we use a different approach (describe idea first, then say that its a generalization).}\tj{Which decoders do you mean? I changed the structure a little. See above.}
The main idea of the algorithm is to randomly guess parts of the error by introducing so-called erasures and trade errors with erasures. This allows to apply an error-erasure decoder (e.g.,~\cite{hoermann2022errorandErasure}) to decode successfully if enough errors were traded with erasures. We analyze the probability of this event for a specific distribution of guessed erasures. The gain comes from the fact, that erasures weigh
% \hb{The weight is not less in terms of the sum-rank metric... the cost is more with respect to the decoding radius.}\tj{,,weigh'' means not weight and also has nothing to do with the sum-rank metric. So I think it's okay to write like this. weigh less w.r.t. decoding capability, not sum-rank weight.}
less than errors with respect to the decoding capability of an \ac{LRS} code.
% In particular, the error-erasure decoder presented in~\cite{hoermann2022errorandErasure} can decode a combination of $t$ errors, $\gamma$ column erasures and $\rho$ row erasures up to
% \begin{equation}
%     2t+\gamma+\rho \leq n-k
% \end{equation}
% where $n$ is the code length and $k$ is the code dimension of the \ac{LRS} code.

Additionally, we demonstrate a method to find the optimal distribution of erasures. 
We show that the proposed algorithm which exploits the structure of the underlying \ac{LRS} code improves over the generic decoding algorithm, introduced in~\cite{puchinger2022generic}, in terms of expected computational complexity.

% \revision{The proofs of all theorems and lemmas appearing in this paper will be available in a future extended version.}

%%%%%%%%%%%%%%%%%%%%%%%%%%%%%%%%%%%%%%%%%%%%%%%%%%%%%%%%%%%%%%%%%%%%%%%%%%%%%%%%%%%
\section{Preliminaries}
\label{sec:preliminaries}

\subsection{Notation}
For a prime power $q$ and a positive integer $m$, let $\Fq$ denote a finite field of order $q$ and $\Fqm$ its extension field of extension degree $m$.
Let $\b=(b_1,\ldots,b_m)\in\Fqm^m$ be a fixed (ordered) basis of $\Fqm$ over $\Fq$. We denote by $\ext(\alpha)$ the column-wise expansion of an element $\alpha\in\Fqm$ over $\Fq$ w.r.t. to the basis $\b$
% \begin{equation*}
    % $\ext : \Fqm \mapsto \Fq^{m}$
% \end{equation*}
s.t. $\alpha = \b \cdot \ext(\alpha)$. This notation is extended to vectors and matrices by applying $\ext(\cdot)$ in an element-wise manner s.t. $\ext : \Fqm^n \mapsto \Fq^{m\times n}$. For a vector $\x\in\Fqm^n$ we define $\rkq(\x) \defeq \rkq(\ext(\x))$.
The $\Fq$-linear row space of a matrix $\B\in\Fq^{m\times n}$ is denoted as $\Rowspace{\B}$
% and the $\Fq$ row space of a vector $\b\in\Fqm^{n}$ is defined as $\$
% The $\Fq$-linear rowspace of a matrix $\A$ over $\Fq$ is denoted by $\Rowspace{\A}$ and its column space by $\Colspace{\A}$.
and the Grassmanian $\Grassm{q}{\mathcal{V}}{i}$ of an $\Fq$-vector space $\mathcal{V}$ is the set of all $i$-dimensional subspaces of $\mathcal{V}$.
% \tj{Not sure if we need this, maybe will be removed but if needed, maybe move to combinatorics subsection of preliminaries?}
\revision{We use the notation $a\sample\mathcal{A}$ to denote an element $a$ drawn uniformly at random from a set $\mathcal{A}$.}

\subsection{Sum-Rank Weight and Linearized Reed--Solomon Codes}

Let $\x=(x_1,\ldots,x_\shots)\in\Fqm^n$ be a vector, that is partitioned into blocks $\x^{(i)}\in\Fqm^{n_i}$ w.r.t. to a \emph{length partition} $\n=(n_1,\ldots,n_\shots)\in\mathbb{N}^\shots$.
The \emph{sum-rank weight} of $\x$ w.r.t. to the length partition $\n$ is then defined as
\begin{equation}
    \SumRankWeight(\x) \defeq \sum_{i=1}^{\shots} \rkq(\x^{(i)}).
\end{equation}
The \emph{sum-rank distance} of two vectors $\x,\y\in\Fqm^n$ is then defined by the sum-rank weight $\SumRankDist(\x,\y)\defeq \SumRankWeight(\x-\y)$.
% \begin{definition}[Linearized Reed--Solomon Code~\cite{martinez2018skew}]
% Let $\conjvec = (\conj_1, \ldots, \conj_\shots) \in \Fqm^\shots$ be a vector containing representatives of pairwise distinct nontrivial conjugacy classes of $\Fqm$ and consider a vector $\n = (n_1, \ldots, n_\shots) \in \mathbb{N}^\shots$ with $n = \sum_{i=0}^{\shots} n_i$.
% Let the vectors $\bm{\beta}^{(i)} = (\beta_1^{(i)},\ldots,\beta_{n_i}^{(i)}) \in \Fqm^{n_i}$ contain $\Fq$-linearly independent elements of $\Fqm$ for all $i=1,\ldots,\shots$ and let $\bm{\beta} = (\bm{\beta}^{(1)} \, | \, \dots \, | \,\bm{\beta}^{(\shots)} ) \in \Fqm^n$.
% an \ac{LRS} code $\LRSC{\automor, \bm{\beta}, \conjvec, \shots; \n, k}\subseteq\Fqm^n$ of length $n$ and dimension $k$ is defined as the set
% \begin{equation}
%     \left\{ (f{(\bm{\beta}^{(1)})}_{\conj_1} \,|\,\dots\,|\, f{(\bm{\beta}^{(\shots)})}_{\conj_\shots}) : f \in \Fqm{[x;\automor]}_{<k}\right\}
% \end{equation}
% where $f{(\bm{\beta}^{(i)})}_{\conj_i} \defeq (f{({\beta_1}^{(i)})}_{\conj_i}, \ldots, f{({\beta_{n_i}}^{(i)})}_{\conj_i} )$ for $i=1,\ldots,\shots$.
% \end{definition}

% \tj{TODO: Make definition of LRS codes as short as possible. Right now would need defintion of skew polynomials, etc. that we can reuse a simple $\LRSCode[q^m; \n,k]$}

\Ac{LRS} codes~\cite{martinez2018skew} are a special class of sum-rank-metric codes which are \ac{MSRD}.
This means that the \emph{minimum sum-rank distance} is $n-k+1$ where $n$ is the code length and $k$ is the code dimension.
Hence, \ac{LRS} codes can uniquely decode errors of weight up to
%\begin{equation}
    $\tau \defeq \left\lfloor \frac{d-1}{2} \right\rfloor$.
%\end{equation}
Throughout this paper we consider \ac{LRS} codes of length $n$, with length partition $\n$ and dimension $k$ over $\Fqm$ which we denote as $\LRSCode$.
Note that \ac{LRS} codes are restricted to $\shots \leq q-1$ and $n_i \leq m$ for all $i=1,\ldots,\shots$ (see~\cite{martinez2018skew}).
%$\shots \leq q-1$ and  $n_i \leq m$ for all $i=1,\ldots,\shots$.
% \begin{itemize}
    % \item $\shots \leq q-1$,
    % \item $n_i \leq m$ for all $i=1,\ldots,\shots$.
% \end{itemize}

\subsection{Channel Model}
Let $\c\in \LRSCode$ and let $\c$ be corrupted by an error $\e$ of sum-rank weight $\errWeight$, i.e.,
the received word is
% \begin{equation}
    $\y = \c + \e$.
% \end{equation}

Any error $\e = (\e^{(1)}\,|\,\dots\,|\,\e^{(\shots)}) \in \Fqm^n$ with $\SumRankWeight(\e) = w$ can be decomposed into a vector-matrix product as
\begin{equation}\label{eq:errorDecomp}
    \e = \a\B %(\a^{(1)}\,|\,\dots\,|\,\a^{(\shots)}) \cdot \diag(\B^{(1)}, \ldots,\B^{(\shots)})
\end{equation}
with $\a \defeq (\a^{(1)}\,|\,\dots\,|\,\a^{(\shots)})$ and $\B \defeq  \diag(\B^{(1)}, \ldots,\B^{(\shots)})$ and with $\a^{(i)}\in\Fqm^{\errWeight_i}$ and $\B\in\Fq^{\errWeight_i \times n_i}$ s.t.
$\rkq(\a^{(i)})=\rkq(\B^{(i)})=\errWeight_i$ and $w = \sum_{i=1}^{\shots} \errWeight_i$ for all $i=1,\ldots,\shots$.
It follows that $\e^{(i)} = \a^{(i)}\B^{(i)}$ for $i=1,\ldots,\shots$ and we have that the entries of $\a^{(i)}$ form a basis of the column space of $\e^{(i)}$ and the rows of $\B^{(i)}$ form a basis of its row space.
% We denote the column space of $\e^{(i)}$ as $\CS^{(i)}$ and the row space as $\RS^{(i)}$ for $i=1,\ldots,\shots$.

The error $\e$ can be further decomposed into a sum of three types of error vectors, namely $\e_{F}$, $\e_{R}$ and $\e_{C}$ s.t. 
\begin{equation}
    \e = \e_{F} + \e_{R} + \e_{C}
\end{equation}
with $\SumRankWeight(\e_{F}) = t$, $\SumRankWeight(\e_{R}) = \rho$ and $\SumRankWeight(\e_{C}) = \gamma$, respectively~\cite{hoermann2022errorandErasure}.
Each of the three vectors can then be decomposed again as in~\eqref{eq:errorDecomp}, with $\a_{F}$, $\B_{F}$, $\a_{R}$, $\B_{R}$ and $\a_{C}$, $\B_{C}$, respectively.
% Note that $\B_{F}$, $\B_{R}$ and $\B_{C}$ also have a block diagonal structure as $\B$ in~\eqref{eq:errorDecomp}.
Assuming neither $\a_{F}$ nor $\B_{F}$ are known, the term $\a_{F}\B_{F}$ is called \emph{full rank errors}. If $\a_{R}$ is known but $\B_{R}$ is unknown, the vector $\a_{R}\B_{R}$ is called \emph{row erasures} and assuming 
$\a_{C}$ is unknown but $\B_{C}$ is known the product $\a_{C}\B_{C}$ is called \emph{column erasures}.

An efficient algorithm for \ac{LRS} codes has been proposed in~\cite{hoermann2022errorandErasure} that is able to correct a combination of \emph{full rank errors}, \emph{row erasures} and \emph{column erasures} up to
\begin{equation}\label{eq:deccond}
    2t + \gamma + \rho \leq n-k
\end{equation}
with a complexity of $\oh{n^2}$ operations over $\Fqm$.
We denote by $\DEC(\y,\a_R,\B_C)$ such an error-erasure decoder that takes as input the received word $\y=\c+\e$, and a basis $\a_R$  of parts of the column spaces (row erasures) and a basis $\B_C$ for parts of the row spaces (column erasures) of the error $\e$ and outputs a valid codeword $\hat{c}$ if~\eqref{eq:deccond} is fulfilled or returns $\emptyset$ else.

\begin{definition}[Row and Column Support]
Let $\e\in\Fqm^n$ be of sum-rank weight $\errWeight$. 
\begin{itemize}
    \item \textbf{Row Support:} The row support of $\e$ is defined as
    \begin{equation}
        \RSv \defeq \RS^{(1)} \times \RS^{(2)} \times \cdots \times \RS^{(\shots)},
    \end{equation}
    where $\RS^{(i)} \subseteq \Fq^{n_i}$ is the $\Fq$-row space of $\B^{(i)}\in\Fq^{\errWeight_i \times n_i}$ and thus of $\e^{(i)}$ as in~\eqref{eq:errorDecomp} for all $i=1,\ldots,\shots$.
    \item \textbf{Column Support:} The column support of $\e$ is defined as
    \begin{equation}
        \CSv \defeq \CS^{(1)} \times \CS^{(2)} \times \cdots \times \CS^{(\shots)},
    \end{equation}
    where $\CS^{(i)} \subseteq \Fq^{m}$ is the column space of $\ext{(\a^{(i)})}\in\Fq^{m\times \errWeight_i}$ and thus of \revision{$\ext{(\e^{(i)})}$} as in~\eqref{eq:errorDecomp} for all $i=1,\ldots,\shots$.
\end{itemize}    
\end{definition}

Assume $\ESv$ to be either a row or column support of the error.
We denote by $\sumDim(\ESv)$ the \emph{sum dimension} of an error support: % which is defined as
\begin{equation}
    \sumDim(\ESv) \defeq \sum_{i=1}^{\shots} \dim(\ES^{(i)}).
\end{equation}
% The sum dimension of the column support is defined analogously for the column support case.
The intersection of two supports $\ES_1$ and $\ES_2$ is defined as
\begin{equation}
    % \ES_1 \cap \ES_2 \defeq \ES_1^{(1)} \cap \ES_2^{(1)} \times \ES_1^{(2)} \cap \ES_2^{(2)} \times \cdots \times \ES_1^{(\shots)} \cap \ES_2^{(\shots)}
    \ESv_1 \cap \ESv_2 \defeq \ES_1^{(1)} \cap \ES_2^{(1)} \times \cdots \times \ES_1^{(\shots)} \cap \ES_2^{(\shots)}.
\end{equation}
% and the dimension of the intersection of the two supports is then simply
% \begin{equation}
%     \sumDim(\ESv_1\cap\ESv_2) = \sum_{i=1}^{\shots} \dim(\ES_1^{(i)} \cap \ES_2^{(i)}).
% \end{equation}

\subsection{Combinatorics}

\begin{definition}[Weight Composition]
Let $\errWeight$, $\shots$ and $\mu$ be non-negative integers s.t. $\errWeight \leq \shots \mu$. We define the set
\begin{equation}
    \wdecomp{\errWeight}{\shots}{\mu} = \left\{ \errWeightVec \in \{0,\ldots,\mu\}^\shots : \sum_{i=1}^{\shots} \errWeight_i = \errWeight \right\},
\end{equation}
which contains all possible weight compositions of a vector consisting of $\shots$ blocks and sum-rank weight $\errWeight$.
This notion is also known in combinatorics as \textbf{weak integer composition}.
\end{definition}

\begin{definition}
    Let $\mu$ be a positive integer and $0\leq s \leq \shots\mu$. For $\s\in\wdecomp{s}{\shots}{\mu}$, we define the set of all supports as
    \begin{equation}
        \Xi_{\mu}(\s) \defeq \left\{ \mathcal{F}_1 \times \cdots \times \mathcal{F}_{\shots} : \mathcal{F}_i \subseteq \Fq^\mu \;\text{s.t.}\, \dim(\mathcal{F}_i)=s_i \right\}.
    \end{equation}
\end{definition}

The number of all matrices in $\Fq^{a\times b}$ with $\Fq$-rank $j$ is
\begin{equation}
    \NM_q(a,b,j) \defeq \quadbinom{a}{j}\cdot\prod_{i=0}^{j - 1}(q^b-q^i)
\end{equation}
where $\quadbinom{a}{j}$ is the Gaussian binomial coefficient defined as
\begin{equation}
    \quadbinom{a}{j} \defeq \prod_{i=1}^{j} \frac{q^{a-j+i}-1}{q^i-1}.
\end{equation}
% \hb{Add reference}

Let $\n\in\mathbb{N}^{\shots}$ be a length partition and let $\mu$ and $\errWeight$ be non-negative integers s.t. $\errWeight\leq\shots\mu$, we denote by $\NSRM{q}{m}{\n}{j}$ the number of vectors in $\Fqm^{n}$ of sum-rank weight exactly $j$.
It is easy to see that we have
\begin{equation}
    \NSRM{q}{m}{\n}{j} = \sum_{\errWeight\in\wdecomp{\errWeight}{\shots}{\mu}} \prod_{i=1}^{\shots} \NM_q(m,n_i, \errWeight_i).
\end{equation}
Note that there are efficient ways to compute $\NSRM{q}{m}{\n}{j}$ (cf.~\cite{puchinger2022generic}).
% If $n_1=\ldots,n_\shots=\eta$ are constant for all $\shots$ blocks, we may write $\NSRM{q}{m}{\eta}{j}$ instead \tj{check if this notation is needed later}.

% \begin{definition}[Integer Partition]
%     Let $t$ and $s$ be positive integers. 
%     An \emph{integer partition} of $t$ into $s$ parts is an $s$-tuple $\bm{\lambda} = (\lambda_1,\ldots,\lambda_s)$ of positive integers with the following properties:
%     \begin{itemize}
%         \item $\sum_{i=1}^{s} \lambda_i = t$,
%         \item $\lambda_1 \geq \lambda_2 \geq \dots \geq \lambda_s$.
%     \end{itemize}
%     For convenience we denote by $\shots_{\bm{\lambda}} = s$ the length of a given integer partition $\bm{\lambda}$.
%     Furthermore, we denote by $\Pi_{\bm{\lambda}}$ the set of all permutations for a given integer partition $\bm{\lambda}$.
%     Let $N_i$ denote the number of occurrences of a positive integer $i$ in a given integer partition $\bm{\lambda}$ of length $s$ and with $0 < i \leq \mu$, then $|\Pi_{\bm{\lambda}}| = \binom{s}{N_1,\dots,N_\mu} = \frac{s!}{N_1 ! \dots N_\mu !}$.
%     By $\IPS_{t,s,\mu}$ we denote the set of all possible integer partitions of $t$ with maximal length $s$ and containing integers with maximum values $\mu$ only.
% \end{definition}

% \begin{equation}
    % \NSRM{q}{m}{\n}{j}
% \end{equation}

% \begin{equation}
%     \e = (\a^{(1)}\,|\,\dots\,|\,\a^{(\shots)}) \cdot \begin{pmatrix}\B^{(1)} & & & \\
%     & \ddots & \\
%     & & \B^{(\shots)}
%     \end{pmatrix}
% \end{equation}

%%%%%%%%%%%%%%%%%%%%%%%%%%%%%%%%%%%%%%%%%%%%%%%%%%%%%%%%%%%%%%%%%%%%%%%%%%%%%%%%%%%
\section{Generic Decoding}
% \hb{Add an introductory sentence here, since otherwise the problem is a bit lost.}\tj{Don't really have space for that imo and I think it's okay. I shifted one sentence infront of the problem. Maybe enough?}

A generic \ac{SRMD} algorithm is an algorithm solving Problem~\ref{prob:searchsrsd}.
\begin{problem}[Search-\acs{SRMD}]
\label{prob:searchsrsd}
\hfill
\begin{itemize}
    \item \textbf{Instance:} Linear sum-rank metric code $\mycode{C} \subseteq \Fqm^n$, $\y\in\Fqm^n$ and an integer $\decRadius > 0$.
    \item \textbf{Objective:} Find a codeword $\c \in \mycode{C}$, s.t. $\SumRankWeight(\y-\c) \leq \decRadius$.
\end{itemize}
\end{problem}
\revision{For $\decRadius \leq \uDR  = \left\lfloor \frac{d-1}{2} \right\rfloor$ at most one solution to Problem~\ref{prob:searchsrsd} exists.}
%For unique decoding there exists a unique solution, i.e. $\decRadius \leq \uDR$, \revision{where $\decRadius$ is the maximum allowed error weight used in Problem~\ref{prob:searchsrsd} and $\tau$ is the unique decoding radius of $\mycode{C}$}.
In general, for decoding beyond the unique decoding radius there might be many solutions to Problem~\ref{prob:searchsrsd} but for our consideration, and as stated in Problem~\ref{prob:searchsrsd}, it is sufficient to find one of them.
A generic decoder for sum-rank metric codes for Problem~\ref{prob:searchsrsd} with $\decRadius \leq \uDR$ is presented in~\cite{puchinger2022generic}
and several bounds on the \revision{computational complexity} of the proposed algorithm are given. 
% \revision{In this paper we refer to the computational complexity as work factor.}
We denote the lower and upper bound on the work factor of~\cite{puchinger2022generic} for solving Problem~\ref{prob:searchsrsd} as $\lbWgen$ and $\ubWgen$, respectively. 
Note that in contrast to~\cite{puchinger2022generic} we consider a constant complexity of a single iteration for the lower bound.

\begin{problem}[Search-\acs{LRS}]
\label{prob:searchlrs}
\hfill
\begin{itemize}
    \item \textbf{Instance:} Linearized Reed--Solomon code $\LRSCode \subseteq \Fqm^n$, $\y\in\Fqm^n$ and an integer $\decRadius > 0$.
    \item \textbf{Objective:} Find $\c \in \LRSCode$, s.t. $\SumRankWeight(\y-\c) \leq \decRadius$.
\end{itemize}
\end{problem}
Problem~\ref{prob:searchlrs} is a special instance of Problem~\ref{prob:searchsrsd}, where the linear code is an \ac{LRS} code. 
% \hb{...and an efficient decoder (i.e. the code structure) is known.}\tj{I don't think there is an efficient decoder for LRS codes beyond the unique decoding radius yet, or what exactly do you mean? I think its okay as is cause in the next section this will be adressed more.}
Currently, the generic decoder from~\cite{puchinger2022generic} has the smallest known complexity to solve Problem~\ref{prob:searchlrs}. 
% \textcolor{red}{As for Gabidulin codes, we should also be able to use a unique decoder (basis for the solution space) and brute-force the rest --- if not we should at least comment on this}
In this paper we show how to reduce the complexity of solving Problem~\ref{prob:searchlrs} compared to the generic decoder \revision{for errors with weight $\errWeight=\SumRankWeight(\y-\c)$, s.t.} $\errWeight > \uDR$. 
In particular, we assume that the excess of the error over the unique decoding radius $\tau$ is larger than zero, i.e. \revision{$\errWeight - \tau > 0$}.
% \begin{equation}
    % \errEx \defeq \tau - \errWeight > 0.
% \end{equation}
%%%%%%%%%%%%%%%%%%%%%%%%%%%%%%%%%%%%%%%%%%%%%%%%%%%%%%%%%%%%%%%%%%%%%%%%%%%%%%%%%%%
\section{Randomized Decoding}\label{sec:randdec}
% In the first part of this section, we analyze the decoding problem under the assumption of having only one codeword in distance $\errWeight$ from the received word $\y$, i.e. $\SumRankWeight(\y-\c)=\errWeight$ and in Subsection~\ref{sec:subsecC} we consider the scenario of $\SumRankWeight(\y-\c)\leq\errWeight$ without the assumption of having exactly one codeword around the decoding radius.
% In this section we propose an algorithm for solving Problem~\ref{prob:searchlrs} for \ac{LRS} codes.
The proposed approach is a generalization of the \emph{randomized decoding algorithm} presented in~\cite{renner2020randomized} from Gabidulin codes (rank metric) to \ac{LRS} codes (sum-rank metric).
In the considered problem we assume an error $\e$ of weight $\SumRankWeight(\e) = w > \tau$ with row support $\RSv$ and column support $\CSv$.
% Since we do not have any knowledge about any of the row spaces $\RS^{(i)}$ or column spaces $\CS^{(i)}$ for any $i=1,\ldots,\shots$
The main idea is to guess parts of the error support.
% \hb{The statement is correct, however, it might help the reader that we guess the spaces in a per-block manner.}\tj{If you read on, this is how it is explained.}
This is done by first drawing a weight composition $\u=(u_1,\ldots,u_\shots)\in\mathbb{N}^\shots$ of the guessed error support, according to a \ac{PMF} $p_{\gsWeightVec}$ and then a guessed support is drawn uniformly at random from $ \Xi_{\mu}(\gsWeightVec)$.
This means, that for each block a row and/or column space is guessed independently with dimension $u_i$ for $i=1,\ldots,\shots$.
If the sum of the dimensions of the intersections of the guessed spaces with the spaces of the actual error is large enough (that means enough errors were traded for erasures) an error-erasure decoder can decode successfully.

In~\cite{renner2020randomized} for Gabidulin codes it was shown, that the algorithm cannot be improved by guessing a combination of row spaces and column spaces.
Therefore, we restrict to guessing only parts of the row support.
Let $\GSv$ with $u\defeq\sumDim(\GSv)$ be the guessed row support then $\gamma = \gsWeight$ and $\rho = 0$. The corresponding weight composition of $\GSv$ is $\u$.
For simplicity, we restrict to the case of \ac{LRS} codes with constant block sizes, that means that $n_1=n_2=\dots=n_\shots$ and we denote $\eta=n/\shots$ for all $i=1,\ldots,\shots$.
In this case the maximum rank of a single block is at most $\mu \defeq \min\{\eta, m\}$.
The adaptation to variable block sizes is straightforward.
The proposed algorithm for guessing only row spaces is presented in Algorithm~\ref{alg:er_aided_rand_dec}.
% \hb{Is there a section title missing? Sounds like we start a new section.}

In this section we analyze an upper bound on the expected number of operations over $\Fqm$ of Algorithm~\ref{alg:er_aided_rand_dec} which solves Problem~\ref{prob:searchlrs}.
% but with the assumption, that there is always at least one codeword around the received word $\y$ within distance $\errWeight$.
Further, we show a method how to evaluate this bound without knowing the actual distribution $p_{\gsWeightVec}$.
A lower bound on the expected number of operations $\Fqm$ is given in Section~\ref{sec:subsecC}.
% which considers the more general case where we assume $\y$ drawn uniformly at random from $\Fqm^n$ is given in Section~\ref{sec:subsecC}. 
Finally, we present a method to compute the optimal distribution $p_{\gsWeightVec}$ that minimizes the worst-case number of iterations of the proposed algorithm.

\begin{algorithm}
    \setstretch{1.35}
    \caption{Randomized \acs{LRS} Decoder}\label{alg:er_aided_rand_dec}
    \SetKwInOut{Input}{Input}\SetKwInOut{Output}{Output}

    \Input{
    Parameters $q, m, k, \n, \shots, \errWeight, \gsWeight$ \\
    Received word $\y\in\Fqm^n$ \\
    LRS error-erasure decoder $\DEC(\cdot,\cdot,\cdot)$\\
    % Erasure weight $\gsWeight$
    % maximum number of iterations $N_\text{max}$
    }

    \Output{$\hat{\c} \in \Fqm^n : \SumRankWeight(\y-\hat{\c}) \leq w$} % or $\emptyset$ (failure)}
    \BlankLine
    % \For{$N=[1, N_\text{max}]$}{
    $\mu \gets \min\{\eta, m\}$ \\
    \While{\KwSty{True}}{
        Draw $\gsWeightVec \in \wdecomp{\gsWeight}{\shots}{\mu}$ according to the distribution $p_{\gsWeightVec}$ \\\label{line:DrawU}
        $\GS \sample \Xi_{\mu}(\gsWeightVec)$ \\\label{line:DrawUII}
         % $\GS^{(i)} \sample \Grassm{q}{\Fq^{n_i}}{\gsWeight_i}$ for all $i=1,\ldots,\shots$ \\
        \For{$j=1,\ldots,\shots$}{
            $\B^{(j)} \gets$ full-rank matrix in $\Fq^{u_j \times \eta}$ s.t $\Rowspace{\B^{(j)}}=\GS_j$ with $\dim(\mathcal{U}_j)=u_j$
        }
        $\B = \diag(\B^{(i)},\ldots,\B^{(\shots)})$ \\
        $\hat{\c} \gets \text{DEC}(\y, \bm{0}, \B)$ \\\label{line:EEDec}
        \If{$\hat{\c}\neq\emptyset$ \KwSty{and} $\SumRankWeight(\y-\hat{\c}) \leq w$}{
             \Return{$\hat{\c}$}
         }
    }
%     \Return{$\hat{\c}$}
\end{algorithm}

\subsection{Upper Bound on the Work Factor of the Algorithm}
Define by $\sdInter$ the sum dimension of the guessed error support $\GSv$ and the actual error support $\ESv$, i.e.
% \begin{equation}
    $\sdInter \defeq \sumDim(\GSv \cap \ESv)$.
% \end{equation}
This means if $\sdInter$ is large enough, we trade errors for erasures and the decoding
condition for an error-erasure decoder such as in~\cite{hoermann2022errorandErasure} is
% \begin{equation}\label{eq:deltaBounds}
    $2(\errWeight-\sdInter)+\gsWeight \leq n-k$
    % \end{equation}
which implies that we should have % $\sdInter \geq \minEps$ with
\begin{equation}\label{eq:minEpsDef}
    % \minEps \defeq \left\lceil \errWeight + \frac{\gsWeight - (n-k)}{2} \right\rceil.
    \sdInter \geq \left\lceil \errWeight + \frac{\gsWeight - (n-k)}{2} \right\rceil.
\end{equation}

\begin{lemma}{\cite[Lemma 1]{renner2020randomized}}\label{lem:interProb}
    Let the error space $\ES^{(i)}$ of the $i$-th component of the error support $\ES$ with $\dim(\ES^{(i)})=\errWeight_i$ and $\gsWeight_i$, the dimension of the $i$-th component $\GS^{(i)}$ of the guessed support be given. Choose $\GS^{(i)}$ uniformly at random from $\Grassm{q}{\Fq^\mu}{\gsWeight_i}$ for all $i=1,\ldots,\shots$. Then, the conditional probability $p_{\errWeight_i,\gsWeight_i}^{(i)}(j) \defeq \Pr[\dim(\mathcal{E}_i \cap \GS_i) = j | \ES^{(i)}, \gsWeight_i]$, that the intersection of $\ES^{(i)}$ and $\GS^{(i)}$ is exactly $j$ is
    \begin{equation}
        p_{\errWeight_i,\gsWeight_i}^{(i)}(j) \defeq \frac{\quadbinom{\mu-\errWeight_i}{\gsWeight_i - j}\quadbinom{\errWeight_i}{j} q^{(\errWeight_i-j)(\gsWeight_i-j)}}{\quadbinom{\mu}{\gsWeight_i}}.
    \end{equation}
\end{lemma}
% \begin{IEEEproof}
% See~\cite[Lemma 1]{renner2020randomized}.
% \end{IEEEproof}
% \tj{Maybe formulate this lemma without the indices?}\textcolor{red}{Without indices it is exactly the same as in [6]?}\tj{Yes}
\begin{lemma}
\label{lem:Sprob}
    Let $\mu$ be a non-negative integer.
    For a fixed error $\e\in\Fqm^n$ and given the weight composition $\gsWeightVec=(\gsWeight_i,\ldots,u_\shots)$ of the guessed space $\GS$ with $\dim(\GS)=u$, choose $\GS$ uniformly at random from $\Xi_{\mu}(\gsWeightVec)$.
    Further let $S_{j}$ be the event that $\sumDim(\ES \cap \GS) = j$. 
    The probability of $S_{j}$ conditioned on $\e$ and $\gsWeightVec$ is then
    \begin{equation}
        \Pr[ S_{j} | \e, \gsWeightVec] =  \left(\,\Conv_{i=1}^{\shots} p_{\errWeight_i,\gsWeight_i}^{(i)} \right)(j)
    \end{equation}
    with
    \begin{equation}
     \left(\,\Conv_{i=1}^{\shots} p_{\errWeight_i,\gsWeight_i}^{(i)}\right)(j) \defeq \left( p_{\errWeight_1,u_1}^{(1)} \conv \cdots \conv p_{\errWeight_\shots,u_\shots}^{(\shots)} \right) (j)
     % (\Conv_{i=1}^{\shots} p_{\ES^{(i)},\gsWeight_i}^{(i)})(j) \defe \left( p_{\ES^{(1)},u_1}^{(1)} \conv p_{\ES^{(2)},u_2}^{(2)} \conv \cdots \conv p_{\ES^{(\shots)},u_\shots}^{(\shots)} \right) (j)
    \end{equation}
    being the $\shots$-fold discrete convolution of the \acp{PMF} $p_{\errWeight_i,\gsWeight_i}^{(i)}$ evaluated at $j$ for all $i=1,\ldots,\shots$.
\end{lemma}
\begin{IEEEproof}
Given the error $\e$ with weight composition $\w$ and given the weight composition $\u$ of the guessed support, let $V_i$ be a random variable that corresponds to the rank of the intersection of the $i$-th guessed space $\GS^{(i)}$ with the $i$-th actual error space $\ES^{(i)}$ for $i=1,\ldots,\shots$.
By Lemma~\ref{lem:interProb} we have that $p_{\errWeight_i,\gsWeight_i}^{(i)}(j) $ is the \ac{PMF} of that event, i.e.
\begin{equation}
 \Pr[V_i = j| \e, \gsWeightVec]=p_{\errWeight_i,\gsWeight_i}^{(i)}(j).   
\end{equation}
Since we are interested in the sum of random variables, i.e. $V = \sum_{i=1}^{\shots}V_i$ the resulting \ac{PMF} is given by the $\shots$-fold discrete convolution of the \acp{PMF} of the random variables $V_i$ for $i=1,\ldots,\shots$. Thus
\begin{equation}
\Pr[V = j| \e, \gsWeightVec] = \left(\,\Conv_{i=1}^{\shots} p_{\errWeight_i,\gsWeight_i}^{(i)}\right)(j)  
\end{equation}
with 
\begin{equation}
    \left(\, p_{\errWeight_1,\gsWeight_1}^{(1)} \Conv p_{\errWeight_2,\gsWeight_2}^{(2)}\right)(j) \defeq \sum_{r=-\infty}^{\infty} p_{\errWeight_1,\gsWeight_1}^{(1)}(r) \, p_{\errWeight_2,\gsWeight_1}^{(2)}(j-r).
\end{equation}
Finally we have that $S_j$ is the event that $V = j$ and this proves the claim.
\end{IEEEproof}
For a given weight composition \revision{$\errWeightVec\in\wdecomp{\errWeight}{\shots}{\mu}$} of the error vector $\e$, each block $\e^{(i)}$ is drawn uniformly at random for $i=1,\ldots,\shots$ and we have that
% \begin{equation}\label{eq:errCondSame}
    $\Pr[S_{j} | \e, \gsWeightVec] = \Pr[S_{j} | \errWeightVec, \gsWeightVec]$
% \end{equation}
for any non-negative integer $j$.

For further analysis we consider the worst-case expected number of iterations of Algorithm~\ref{alg:er_aided_rand_dec} until an appropriate guessed error support $\GS$ is drawn s.t. the
error-erasure decoder can successfully decode.
For given weight compositions $\errWeightVec\in\wdecomp{\errWeight}{\shots}{\mu}$ and $\gsWeightVec\in\wdecomp{\gsWeight}{\shots}{\mu}$ of the actual error and the guessed spaces, respectively, we define
\begin{equation}
 \prob_{\mu}(\gsWeightVec,\errWeightVec) \defeq \sum_{j=\left\lceil\errWeight+\frac{\gsWeight-(n-k)}{2}\right\rceil}^{\min[u,\errWeight]}\Pr[S_{j} | \gsWeightVec, \errWeightVec]  
\end{equation}
and for a given \ac{PMF} $p_{\gsWeightVec}$ of $\gsWeightVec$ we have
\begin{equation}\label{eq:errPgivenU}
    \prob_{\mu,\gsWeight}(\errWeightVec) \defeq \sum_{\gsWeightVec\in\wdecomp{\gsWeight}{\shots}{\mu}}p_{\gsWeightVec}\prob_{\mu}(\gsWeightVec,\errWeightVec).
\end{equation}
The worst-case probability $\prob_{\mu, \gsWeight}(\errWeight)$ that maximizes the number of iterations over all possible weight compositions $\errWeightVec$ is
\begin{equation}\label{eq:minErrP}
    \prob_{\mu,\gsWeight}(\errWeight) \defeq \min_{\errWeightVec\in\wdecomp{\errWeight}{\shots}{\mu}} \prob_{\mu,\gsWeight}(\errWeightVec)
\end{equation}
which implies that
\begin{equation}\label{eq:expIter}
    \max_{\errWeightVec\in\wdecomp{\errWeight}{\shots}{\mu}} \mathbb{E}[\text{\#iterations}] = {\prob_{\mu,\gsWeight}(\errWeight)}^{-1}.
    % \max_{\errWeightVec\in\wdecomp{\errWeight}{\shots}{\mu}} \left\{ \frac{1}{\prob_{\mu,\gsWeight}(\errWeightVec)} \right\}
\end{equation}

\begin{theorem}\label{the:wcWF}
Let $\y = \c + \e$ with $\c \in \LRSCode$ and $\errWeight = \SumRankWeight(\e) > \tau$. 
Then, Algorithm~\ref{alg:er_aided_rand_dec} with input $\y$ returns $\hat{\c} \in \LRSCode$ s.t. $\SumRankWeight(\y-\hat{\c})=\errWeight$
and the expected number of operations over $\Fqm$ to output $\hat{\c}\in\LRSCode$ for $\gsWeight=\sumDim{(\GS)}$ is at most
\begin{equation}\label{eq:Wran}
    \WrandUB = \frac{n^2 \shots^\gsWeight}{\prob_{\mu,\gsWeight}(\errWeight) }
\end{equation}
 with $\prob_{\mu,\gsWeight}(\errWeight)$ as in~\eqref{eq:minErrP}.    
\end{theorem}
\begin{IEEEproof}
Drawing from the distribution $p_{\gsWeightVec}$ in Line~\ref{line:DrawU} draws from the set $\wdecomp{\gsWeight}{\shots}{\mu}$ which according to the bound introduced in~\cite{puchinger2022generic} has cardinality at most $|\wdecomp{\gsWeight}{\shots}{\mu}| \leq \binom{\shots+\gsWeight-1}{\shots-1}$. For a fixed $\gsWeight$, this means that the set size $|\wdecomp{\gsWeight}{\shots}{\mu}|$ is in $\oh{\shots^\gsWeight}$.
One iteration of Algorithm~\ref{alg:er_aided_rand_dec} (Line~\ref{line:EEDec}) costs $\oh{n^2}$ operations over $\Fqm$ (see~\cite{hoermann2022errorandErasure}).
This means in total we have a complexity of $n^2 \shots^\gsWeight$ for a single iteration in Algorithm~\ref{alg:er_aided_rand_dec}.
Since we have that $\y = \c + \e$ with $\SumRankWeight(\e)=\errWeight$ we know that there is at least one valid codeword s.t. $\SumRankWeight(\y-\hat{\c})\leq\errWeight$ and since $\prob_{\mu,\gsWeight}(\errWeight)$ is by definition the smallest probability over all $\errWeightVec$ for the algorithm to succeed, we have that the expected number of iterations is at most $\WrandUB$ as in~\eqref{eq:Wran}.
\end{IEEEproof}
In order to evaluate $\WrandUB$, the \ac{PMF} $p_{\gsWeightVec}$ must be known for $\gsWeightVec\in\wdecomp{\gsWeight}{\shots}{\mu}$.
Theorem~\ref{the:Wbounds} gives a lower and upper bound on $\WrandUB$ which both do not depend on $p_{\gsWeightVec}$.
\begin{theorem} \label{the:Wbounds}
% Let $\y = \c + \e$ with $\c \in \LRSC{q^m; n, k}$ and $\SumRankWeight(\e) = \errWeight$. 
% Then, Algorithm~\ref{alg:er_aided_rand_dec} with input $\y$ returns a codeword $\hat{\c}\c \in \LRSC{q^m; n, k}$ s.t. $\SumRankWeight(\y-\c)=\errWeight$.
Let the same conditions hold as in Theorem~\ref{the:wcWF}
The work factor $\WrandUB$ on the expected complexity can then be bounded from below and above as follows:
% operations over $.
\begin{equation}
  \lbWrand  \leq \WrandUB \leq \ubWrand
\end{equation}
with
\begin{equation}
    \lbWrand = n^2 \shots^\gsWeight \cdot \frac{Q_{\mu,\errWeight,\gsWeight}}{{|\wdecomp{\errWeight}{\shots}{\mu}|}}
\end{equation}
and
\begin{equation}
    \ubWrand = n^2 \shots^\gsWeight \cdot Q_{\mu,\errWeight,\gsWeight}
\end{equation}
where
% \begin{equation}
%     Q_{\mu,\errWeight,\gsWeight} \defeq \sum_{\errWeightVec \in \IPS_{\errWeight,\shots,\mu}} \frac{|\Pi_{\errWeightVec}|\binom{\shots}{\shots_{\errWeightVec}}}{\max\limits_{\gsWeightVec\in\wdecomp{\gsWeight}{\shots_{\errWeightVec}}{\mu}}\prob_{\mu}(\gsWeightVec, \errWeightVec)}.
% \end{equation}
\begin{equation}
    Q_{\mu,\errWeight,\gsWeight} \defeq \sum_{\errWeightVec \in \wdecomp{\errWeight}{\shots}{\mu}} \frac{1}{\max\limits_{\gsWeightVec\in\wdecomp{\gsWeight}{\shots}{\mu}}{\prob_{\mu}(\gsWeightVec, \errWeightVec)}}.
\end{equation}
\end{theorem}

\begin{IEEEproof}
First, define 
 \begin{equation}
 \hat{\gsWeightVec} = \xi_{\mu,\gsWeight}(\errWeightVec) \defeq \argmax_{\gsWeightVec\in\wdecomp{\gsWeight}{\shots}{\mu}}{\prob_{\mu}(\gsWeightVec, \errWeightVec)}
 \end{equation}
 i.e. $\xi_{\mu,\gsWeight}(\errWeightVec)$ returns the weight composition $\hat{\gsWeightVec}$ that maximizes $\prob_{\mu}(\gsWeightVec, \errWeightVec)$ for a given $\errWeightVec$ over all $\gsWeightVec\in\wdecomp{\gsWeight}{\shots}{\mu}$.
 We have that
 \begin{equation}\label{eq:argmaxMaxMax}
     \prob_{\mu}( \xi_{\mu,\gsWeight}(\errWeightVec), \errWeightVec) = \max_{\gsWeightVec\in\wdecomp{\gsWeight}{\shots}{\mu}}{\prob_{\mu}(\gsWeightVec, \errWeightVec)}.
 \end{equation}
 Consider, that instead of choosing a vector $\gsWeightVec\in\wdecomp{\gsWeight}{\shots}{\mu}$ directly, we draw a vector $\errWeightVec\in\wdecomp{\errWeight}{\shots}{\mu}$ at random according to a designed probability distribution, defined as
 \begin{equation}
 \tilde{p}_{\errWeightVec} \defeq \frac{1}{{\prob_{\mu}( \xi_{\mu,\gsWeight}(\errWeightVec), \errWeightVec)}}\cdot {Q_{\mu,\errWeight,\gsWeight}^{-1}}.
 \end{equation}
 Denote by $\tilde{p}_{\gsWeightVec}$ the resulting probability distribution of $\gsWeightVec$, for a fixed error $\errWeightVec_{e}$. By~\eqref{eq:errPgivenU} we have that
 \begin{eqnarray}
      \prob_{\mu,\gsWeight}({\errWeightVec}_{e}) &=& \sum_{\gsWeightVec\in\wdecomp{\gsWeight}{\shots}{\mu}}\tilde{p}_{\gsWeightVec}\prob_{\mu}(\gsWeightVec,{\errWeightVec}_{e}) \\
      &=& \sum_{\errWeightVec\in\wdecomp{\errWeight}{\shots}{\mu}}\tilde{p}_{\errWeightVec}\prob_{\mu}(\xi_{\mu}(\errWeightVec,\gsWeight),{\errWeightVec}_{e}) \\
      &\geq& \tilde{p}_{\errWeightVec_{e}}\prob_{\mu}(\xi_{\mu}(\errWeightVec_{e},\gsWeight),{\errWeightVec}_{e}) \\
      &=& Q_{\mu,\errWeight,\gsWeight}^{-1}.
\end{eqnarray}
The value of $Q_{\mu,\errWeight,\gsWeight}$ does not depend on $\errWeightVec_{e}$ anymore and thus holds for all $\prob_{\mu,\gsWeight}({\errWeightVec}_{e})$ with any ${\errWeightVec}_{e}\in\wdecomp{\errWeight}{\shots}{\mu}$ and therefore $\prob_{\mu,\gsWeight}(\errWeight) \geq Q_{\mu,\errWeight,\gsWeight}^{-1}$. Considering the same costs of one iteration in Algorithm~\ref{alg:er_aided_rand_dec} as in Theorem~\ref{the:wcWF} proves the upper bound.
By~\eqref{eq:argmaxMaxMax} and assuming that $\errWeightVec_{e}$ is the weight composition of the worst-case error vector that minimizes~\eqref{eq:minErrP} we have that
\begin{eqnarray}
 % \prob_{\mu}(\gsWeight, \errWeight) &=& \min_{\errWeightVec\in\wdecomp{\errWeight}{\shots}{\mu}} \prob_{\mu,\gsWeight}(\errWeightVec) \\
      \prob_{\mu,\gsWeight}(\errWeight) &=& \prob_{\mu,\gsWeight}({\errWeightVec}_{e})\\
       &=& \textstyle\sum_{\errWeightVec\in\wdecomp{\errWeight}{\shots}{\mu}}\tilde{p}_{\errWeightVec}\prob_{\mu}(\xi_{\mu}(\errWeightVec,\gsWeight),{\errWeightVec}_{e}) \\
      &\leq& \textstyle\sum_{\errWeightVec\in\wdecomp{\errWeight}{\shots}{\mu}}\tilde{p}_{\errWeightVec}\prob_{\mu}(\xi_{\mu}(\errWeightVec,\gsWeight),{\errWeightVec}) \\
      &=& \textstyle\sum_{\errWeightVec\in\wdecomp{\errWeight}{\shots}{\mu}} Q_{\mu,\errWeight,\gsWeight}^{-1} = |\wdecomp{\errWeight}{\shots}{\mu}| Q_{\mu,\errWeight,\gsWeight}^{-1},
\end{eqnarray}
which proves the claim for the lower bound.
\end{IEEEproof}

\subsection{Lower Bound on the Work Factor of the Algorithm}\label{sec:subsecC}
In the previous section we obtained an upper bound on the worst-case number of iterations needed for Algorithm~\ref{alg:er_aided_rand_dec} to output a valid codeword $\hat{\c}\in\LRSCode$ s.t. $\SumRankWeight(\y-\hat{\c})=\errWeight$ where we assumed $\y = \c + \e \in\Fqm^n$ with $\SumRankWeight(\e)=\errWeight$. In this setting, there is at least one codeword in distance $\errWeight$ around the received word $\y$. Neither Problem~\ref{prob:searchsrsd} nor Problem~\ref{prob:searchlrs} make any assumptions on the received word $\y$. Since the decoder is limited to a maximum radius $\errWeight$, in general there can be potentially many more solutions to our decoding problem or none at all.
In this section we consider a lower bound on the number of iterations needed for Algorithm~\ref{alg:er_aided_rand_dec} to output a valid codeword $\hat{\c}\in\LRSCode$ s.t. $\SumRankWeight(\y-\hat{\c})\leq\errWeight$ and we assume that $\y$ is drawn uniformly at random from $\Fqm^n$.
\begin{theorem}\label{the:WFUB}
    Let $\y$ be uniformly drawn at random from $\Fqm^n$.
    % \revision{Add here a sentence that $\LRSCode$ is from a code ensemble? Also need to define code ensemble.}
    Then the average work factor of Algorithm~\ref{alg:er_aided_rand_dec} to output $\hat{\c}\in\LRSCode$ s.t. $\SumRankWeight(\y-\hat{\c})\leq \errWeight$ is at least
    % Then, for $u=\gsWeight$ the probability that an \ac{LRS} error-erasure decoder using a random guess for $\GS$ according to Line~\ref{line:DrawU} and Line~\ref{line:DrawUII} in Algorithm~\ref{alg:er_aided_rand_dec} outputs $\c\in\LRSCode[q^m; \n,k]$ s.t. $\SumRankWeight(\y-\c)\leq \errWeight$ is at most
    \begin{equation}
        \WrandLB = \frac{n^2 \shots^\gsWeight}{\sum_{j=0}^{\errWeight} \bar{A}_j \hat{\prob}_{\mu,\gsWeight}(j) }
    \end{equation}
    with
    \begin{equation}\label{eq:avgA}
         \bar{A}_j \defeq q^{m(k-n)} \NSR(m,\n,j)
    \end{equation}
    and 
    \begin{equation}\label{eq:bestcaseP}
    \hat{\prob}_{\mu,\gsWeight}(\errWeight) \defeq \max_{\errWeightVec\in\wdecomp{\errWeight}{\shots}{\mu}} \prob_{\mu,\gsWeight}(\errWeightVec).
    \end{equation}
\end{theorem}
\begin{IEEEproof}
     Let $\hat{\mathcal{C}}$ be the set of codewords that have rank distance at most $\errWeight$ from the received word, i.e.,
    \begin{equation}
        \hat{\mathcal{C}} \defeq \left\{ \c\in\LRSCode : \SumRankWeight(\y-\c) \leq \errWeight \right\} = \{ \hat{\c}_1, \ldots, \hat{\c}_N \}.
    \end{equation}
    Further, let $X_i$ be the event that the error-erasure decoder outputs $\hat{\c}_i$ for any $i=1,\ldots,N$ and let
    \begin{equation}
        \mathcal{A}_j \defeq \{\hat{\c}_i \in\hat{\mathcal{C}} : \SumRankWeight(\y-\hat{\c}_i) = j\}.
    \end{equation}
    The probability of success over all $\y\in\Fqm^n$ is
    \begin{equation}
        \sum_{\y\in\Fqm^n}p_{\y}\Pr\left[ \bigcup\limits_{i=1}^{N} X_i | \y \right] \leq \sum_{\y\in\Fqm^n}p_{\y} \sum_{i=1}^{N}  \Pr[X_i | \y]
    \end{equation}
    with $p_{\y} = {|\Fqm^n|}^{-1} = q^{-mn}$.
    Denote with $\wcomp(\cdot)$ the function that returns the error weight composition $\w\in\mathbb{N}^\shots$ of a given error vector $\e\in\Fqm^n$ s.t. $\w=\wcomp(\e)$.
    We then have that
    \begin{eqnarray*}
        \sum_{\y\in\Fqm^n} \sum_{i=1}^{N} p_{\y} \Pr[X_i | \y] &=& \sum_{\y\in\Fqm^n}\sum_{i=1}^{N} p_{\y} \prob_{\mu,\gsWeight}(\wcomp(\y-\hat{\c}_i)) \\
        &\leq& \sum_{\y\in\Fqm^n}\sum_{i=1}^{N} p_{\y} \hat{\prob}_{\mu,\gsWeight}(\SumRankWeight(\y-\hat{\c}_i)).
    \end{eqnarray*}
    Since $\hat{\prob}_{\mu,\gsWeight}(\SumRankWeight(\y-\hat{\c}_i))$ with $\hat{\prob}_{\mu,\gsWeight}(\cdot)$ as defined in~\eqref{eq:bestcaseP} is the same for all $\hat{c}_i\in\mathcal{A}_{\SumRankWeight(\y-\hat{\c}_i)}$
    we have that
    \begin{eqnarray*}
        \sum_{\y\in\Fqm^n} \sum_{i=1}^{N} p_{\y} \Pr[X_i | \y] &=& \sum_{\y\in\Fqm^n}\sum_{i=0}^{w} p_{\y} \cdot |\mathcal{A}_i| \cdot \hat{\prob}_{\mu,\gsWeight}(i). \\
        % &=& \sum_{i=0}^{w}\hat{\prob}_{\mu,\gsWeight}(i) \sum_{\y\in\Fqm^n}  p_{\y} |\mathcal{A}_i|.
        % &\leq& \sum_{\y\in\Fqm^n}\sum_{i=1}^{N} p_{\y} \hat{\prob}_{\mu,\gsWeight}(\SumRankWeight(\y-\hat{\c}_i)) 
    \end{eqnarray*}
    For the last step, we have that on average it holds that $\sum_{\y\in\Fqm^n}  p_{\y} |\mathcal{A}_i|=\bar{A}_i$ with $\bar{A}_i$ as in~\eqref{eq:avgA} and thus
    \begin{equation}
         \sum_{\y\in\Fqm^n}p_{\y}\Pr\left[ \bigcup\limits_{i=1}^{N} X_i | \y \right] \leq \sum_{i=0}^{w}\hat{\prob}_{\mu,\gsWeight}(i) \bar{A}_i.
    \end{equation}
    The bound for the work factor then follows by considering the complexity of a single iteration divided by the probability of success.
\end{IEEEproof}

\subsection{Finding the Optimal Drawing Distribution}
Similar to~\cite{puchinger2022generic}, the problem of minimizing~\eqref{eq:expIter} over all distributions $p_{\gsWeightVec}$ on $\wdecomp{u}{\shots}{\mu}$ can be formulated as a \emph{linear program} and solved numerically for small parameters $\shots$, $\mu$ and $u$. %\hb{Refer to Sven's paper and say that it's inspired by their approach.}
\begin{theorem}\label{the:LP}
Let $N_{\gsWeight}=|\wdecomp{\gsWeight}{\shots}{\mu}|$ and $N_{\errWeight}=|\wdecomp{\errWeight}{\shots}{\mu}|$ and fix arbitrary orders $\gsWeightVec_{1},\ldots,\gsWeightVec_{N_{\gsWeight}}$ and $\errWeightVec_{1},\ldots,\errWeightVec_{N_{\errWeight}}$ of all elements in $\wdecomp{\gsWeight}{\shots}{\mu}$ and $\wdecomp{\errWeight}{\shots}{\mu}$, respectively. Further, let
    \begin{eqnarray}
        \c &=& {(0,0,\ldots,0,1)}^\top \in \mathbb{R}^{(N_{\gsWeight}+1) \times 1} \\
        \b &=& {(0,0,\ldots,0,1,-1)}^\top \in \mathbb{R}^{(N_{\errWeight}+2) \times 1}
    \end{eqnarray}
    and
    \begin{equation}
        \A = \begin{pmatrix}
        -\prob_{\mu}(\gsWeightVec_{1},\errWeightVec_{1}) & \dots & -\prob_{\mu}(\gsWeightVec_{N_{\gsWeight}},\errWeightVec_{1}) & 1 \\
        \vdots & \ddots & \vdots & \vdots \\
        -\prob_{\mu}(\gsWeightVec_{1},\errWeightVec_{N_{\errWeight}}) & \cdots & -\prob_{\mu}(\gsWeightVec_{N_{\gsWeight}},\errWeightVec_{N_{\errWeight}}) & 1 \\
        1 & \dots & 1 & 0 \\
        -1 & \dots & -1 & 0 \\
        \end{pmatrix}
    \end{equation}
    with $\A\in\mathbb{R}^{(N_{\errWeight}+2) \times (N_{\gsWeight}+1)}$.
    If $\x=(x_1,\ldots,x_{N_{\gsWeight}+1})$ with $\x\in\mathbb{R}^{(N_{\gsWeight}+1)\times 1}$ is a solution to the linear program
    \begin{itemize}
        \item Maximize $\c^\top\x$
        \item subject to $\A\x \leq \b$ and $\x\geq\bm{0}$,
    \end{itemize}
    then $\tilde{p}_{\gsWeightVec} = x_i$, for all $i=1,\ldots,N_{\gsWeight}$, is a distribution that minimizes~\eqref{eq:expIter} and we have
    \begin{equation}\label{eq:xMax}
        x_{N_{\gsWeight}+1} = \max_{\bm{p}_{\gsWeightVec}} {\prob_{\mu,\gsWeight}(\errWeight)}
    \end{equation}
    with ${\prob_{\mu, \gsWeight}(\errWeight)}$ as defined in~\eqref{eq:minErrP} and $\bm{p}_{\gsWeightVec} = (p_{\gsWeightVec_1},\ldots,p_{\gsWeightVec_{N_{\gsWeight}}})$ with ${\bm{p}}_{\gsWeightVec}\in{[0,1]}^{N_{\gsWeight}}$ s.t. $\sum_{i=1}^{N_{\gsWeight}} p_{\gsWeightVec_i}=1$.    
\end{theorem}
\begin{IEEEproof}
    Let $\tilde{p}_{\gsWeightVec_i} = x_i$ then the last two rows of $\A$ and the last two entries of $\b$ correspond to $\sum_{i=1}^{N_{\gsWeight}} \tilde{p}_{\gsWeightVec_i} = 1$. Together with $\x \geq \bm{0}$, we get that $\tilde{p}_{\gsWeightVec_i}$ is a valid \ac{PMF}.
    The first $N_{\errWeight}$ rows of $\A$ correspond to the constraints
    \begin{equation}
        \sum_{i=1}^{N_{\gsWeight}} \tilde{p}_{\gsWeightVec_i}\prob_{\mu}(\gsWeightVec_i, \errWeightVec_j) \geq x_{N_{\gsWeight}+1} \quad \forall j=1,\ldots,N_{\errWeight}.
    \end{equation}
    Since $x_{N_{\gsWeight}+1}$ is the maximal positive value for which this constraint is fulfilled for all $j=1,\ldots,N_{\errWeight}$ and all solutions $\tilde{p}_{\gsWeightVec_i}$, we have
    \begin{equation}
        x_{N_{\gsWeight}+1} = \max_{\bm{p}_{\gsWeightVec}} \left\{\min_{j=1,\ldots,N_{\errWeight}} \left\{\sum_{i=1}^{N_{\gsWeight}} p_{\gsWeightVec_i}\prob_{\mu}(\gsWeightVec_i,\errWeightVec_j) \right\}\right\}
    \end{equation}
    which is equivalent to~\eqref{eq:xMax} due to the definitions in~\eqref{eq:minErrP} and~\eqref{eq:errPgivenU}.
    % \begin{equation}
    %     \min_{j=1,\ldots,N_{\errWeight}} \left\{\sum_{i=1}^{N_{\gsWeight}} p_{\gsWeightVec_i}\prob_{\mu}(\gsWeightVec_i,\errWeightVec_j) \right\} =  \min_{\errWeightVec\in\wdecomp{\errWeight}{\shots}{\mu}} \sum_{\gsWeightVec\in\wdecomp{\gsWeight}{\shots}{\mu}}p_{\gsWeightVec}\prob_{\mu}(\w,\gsWeightVec)
    % \end{equation}
\end{IEEEproof}
The worst-case complexity using the \ac{PMF} $\tilde{p}_{\gsWeightVec}$ obtained via the linear program is then given by
\begin{equation}
    \Wropt \defeq \frac{n^2 \shots^\gsWeight}{x_{N_{\gsWeight}+1}}
\end{equation}
where $n^2 \shots^\gsWeight$ is the approximate cost of a single iteration in Algorithm~\ref{alg:er_aided_rand_dec} as stated in Theorem~\ref{the:wcWF} and ${x_{N_{\gsWeight}+1}}^{-1}$ is the worst-case number of iterations using $\tilde{p}_{\gsWeightVec}$ as stated in Theorem~\ref{the:LP}.

%%%%%%%%%%%%%%%%%%%%%%%%%%%%%%%%%%%%%%%%%%%%%%%%%%%%%%%%%%%%%%%%%%%%%%%%%%%%%%%%%%%
\section{Numerical Results}
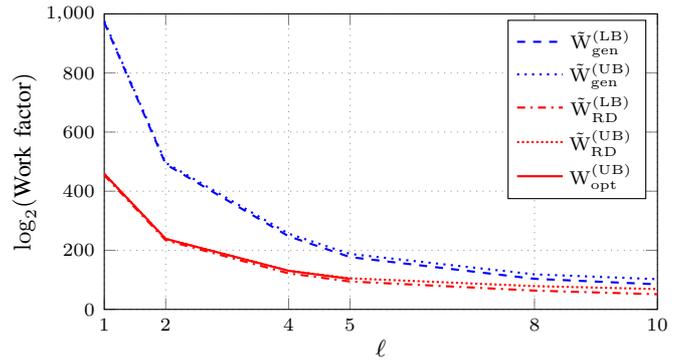
\begin{figure}[ht]
  \begin{center}
    \begin{tikzpicture}
      \begin{axis}[
          width=\linewidth, % Scale the plot to \linewidth
          height=5.5cm,
          grid=major, 
          grid style={dotted,gray!80},
          xlabel={$\shots$}, % Set the labels
          ylabel={$\log_2(\text{Work factor})$},
          xmax=10,
          xmin=1,
          ymax=1000,
          ymin=0,
          xtick={1,2,4,5,8,10},
          tick label style={font=\scriptsize},
          legend style={font=\scriptsize}, 
          label style={inner sep=0, font=\small},
          % legend style={at={(0.5,-0.2)},anchor=north},
          %x tick label style={rotate=90,anchor=east}
        ]
        
        \addplot[line width=0.8pt, color=blue, dashed,mark options=solid] 
        table[x=ell,y=W,col sep=comma] {./results/Wglb_q16_n40_k20_m4_t12_d4.txt}; 
        
        \addplot[line width=0.8pt, color=blue, dotted,mark options=solid] 
        table[x=ell,y=W,col sep=comma] {./results/Wgub_q16_n40_k20_m4_t12_d4.txt}; 
        
        \addplot[line width=0.8pt, color=red, dash dot,mark options=solid]
        table[x=ell,y=W,col sep=comma] {./results/Wrlb_q16_n40_k20_m4_t12_d4.txt}; 
        
        \addplot[line width=0.8pt, color=red, densely dotted, mark options=solid]
        table[x=ell,y=W,col sep=comma] {./results/Wrub_q16_n40_k20_m4_t12_d4.txt}; 

        \addplot[line width=0.8pt, color=red, solid,mark options=solid,]
        table[x=ell,y=W,col sep=comma] {./results/Wropt_q16.txt}; 
        
        \legend{$\lbWgen$,$\ubWgen$,$\lbWrand$,$\ubWrand$,$\Wropt$}    
      \end{axis}
    \end{tikzpicture}
    \vspace{-0.2cm}
    \caption{For \ac{LRS} codes with parameters $q=2^4$, $n=40$, $k=20$, and $m=\eta=n/\shots$, with errors of weight $\errWeight = 12$  we compare the bounds for the generic decoder proposed in~\cite{puchinger2022generic} for $s=20$  and the bounds given in Theorem~\ref{the:Wbounds} with $\gsWeight=4$ for Algorithm~\ref{alg:er_aided_rand_dec}.}\label{fig:randVsGen}
  \end{center}
\end{figure}

In this section we evaluate the tightness of 
%\hb{better: evaluate (the tightness). Otherwise it sounds like we validate the correctness of our results, which is not the case.}
the bounds on the work factor of Algorithm~\ref{alg:er_aided_rand_dec} given in Section~\ref{sec:randdec}.
Figure~\ref{fig:randVsGen} shows the comparison of the bounds on the worst-case number of operations over $\Fqm$ for both the generic decoder from~\cite{puchinger2022generic} and the proposed algorithm with the assumption, that the received word $\y$ is $\y = \c + \e$ with an error $\e$ s.t. $\SumRankWeight(\e)=\errWeight$.
We also give the upper bound $\Wropt$ w.r.t. the worst-case number of iterations derived from the optimal distribution, discussed in Section~\ref{sec:subsecC}.
We observe that the bounds as given in Theorem~\ref{the:Wbounds} which can be computed without any knowledge of the distribution $p_{\gsWeightVec}$ of the weight composition $\gsWeightVec$ of the guessed supports $\GS$ are relatively tight and the work factor for the optimal distribution $\Wropt$ lies in between those bounds.

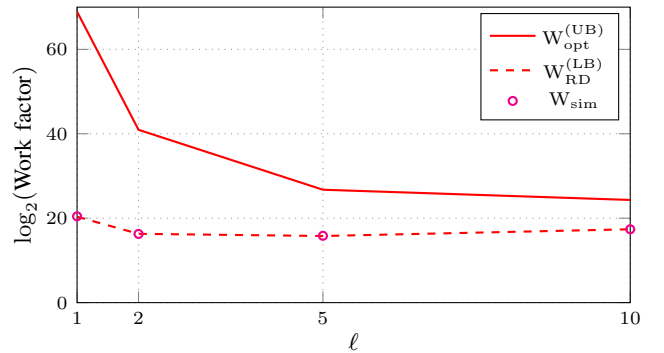
\begin{figure}[ht]
  \begin{center}
    \begin{tikzpicture}
      \begin{axis}[
          width=\linewidth, % Scale the plot to \linewidth
          height=5.5cm,
          grid=major, 
          grid style={dotted,gray!80},
          xlabel={$\shots$}, % Set the labels
          ylabel={$\log_2(\text{Work factor})$},
          xmax=10,
          xmin=1,
          ymax=70,
          ymin=0,
          xtick={1,2,5,10},
          % x unit=\si{\volt}, % Set the respective units
          % y unit=\si{\ampere},
          %legend style={at={(0.5,-0.2)},anchor=north},
          %x tick label style={rotate=90,anchor=east}
          tick label style={font=\scriptsize},
          legend style={font=\scriptsize}, 
          label style={inner sep=0, font=\small},
        ]
        \addplot[line width=0.8pt, color=red, solid,mark options=solid] 
        table[x=ell,y=W,col sep=comma] {./results/Wrub_q11_n10_k5_m1_t4_d3.txt}; 
        \addplot[line width=0.8pt, color=red, dashed,mark options=solid] 
        table[x=ell,y=W,col sep=comma] {./results/WrlbAmod_q11_n10_k5_m1_t4_d3.txt}; 
        \addplot[line width=0.8pt, color=magenta, mark=o,mark options=solid, only marks, mark size=1.5pt] 
        table[x=ell,y=W,col sep=comma] {./results/Wsim_q11.txt}; 
        \legend{$\Wropt$,$\WrandLB$,$\Wsim$}        
      \end{axis}
    \end{tikzpicture}
    \vspace{-0.2cm}
    \caption{Simulation results of Algorithm~\ref{alg:er_aided_rand_dec} for \ac{LRS} codes with small parameters $q=11$, $n=10$, $k=5$, and $m=\eta=n/\shots$, with errors of weight $\errWeight = 4$ with $\gsWeight=3$ for different values of $\shots$.}\label{fig:simulation}
  \end{center}
\end{figure}

Further, we back up the correctness of the lower bound for the scenario that $\y\sample\Fqm^n$ by simulations. The lower bound and the simulation is shown in Figure~\ref{fig:simulation} for small code parameters of $q=11$, $n=10$ and $k=5$.
The simulations were performed using the error-erasure decoder for \ac{LRS} codes from~\cite{hoermann2022errorandErasure} and running for a maximum samples size of $10^7$ vectors $\y$ drawn uniformly at random from $\Fqm^n$.
For reference we also depict the upper bound $\Wropt$ for the optimal distribution obtained from the linear program discussed in Section~\ref{sec:subsecC} as well.

% \begin{equation}
%     \bar{A}_j = q^{m(k-n)}\NSRM{q}{m}{\n}{j}
% \end{equation}

\section{Conclusion}
\label{sec:conclusion}
We presented a randomized decoding algorithm for \ac{LRS} codes that can correct errors beyond the unique decoding radius and analyzed its theoretical expected complexity.
% its theoretical runtime in terms of the expected number of iterations needed to terminate.
We showed that the algorithm improves upon the generic decoding approach from~\cite{puchinger2022generic} by exploiting the structure of the underlying \ac{LRS} code.
% Besides being of theoretical interest, we think that the results given in this paper also open up the possibility to investigate \ac{LRS} codes for code-based cryptography.
The problem of decoding \ac{LRS} codes beyond their unique decoding radius has exponential complexity and thus it can be of interest \revision{to analyze} future code-based cryptosystems in the sum-rank metric that are based on
the hardness of decoding beyond the unique decoding radius.
% \tj{Maybe mention FL system here? Such as FL system in the rank metric for Gabidulin codes~\cite{faure2006new,wachterzeh2018repairing}.} \aw{Yes, that would be good. But call it something like "cryptosystems based on the hardness of list decoding/decoding beyond the unique decoding radius}
% \hb{This sentence sounds a bit vague (same as in abstract). We can say that solving the considered problem has exponential complexity and thus it can be of interest for future code-based cryptosystems in the sum-rank metric.}\tj{Yes I know, I took this kind of argumentation from the generic decoding paper. Also not really sure what else to write tbh}
\revision{Future work will include a more detailed analysis of the bounds on the expected complexity of the algorithm as well as complexity analysis of the evaluation of the bounds.}

% \todo{
% Preliminaries/Notation:
% \begin{itemize}
%     % \item Work factor definition
%     % \item $q$-binomial
%     \item Do we need Grassmanian? Maybe talk about how errors are drawn uniformly?
% \end{itemize}
% }
% \clearpage
\enlargethispage{-0.5cm} 
% \IEEEtriggeratref{12}
% \IEEEtriggercmd{\enlargethispage{-10cm}}
\bibliographystyle{IEEEtran}
\bibliography{references}
%%%%%% 
% Appendix:
% If needed a single appendix is created by
%
% \clearpage
% \appendices
%%
%% If several appendices are needed, then the command
%%
% \appendices
%%
%% in combination with further \section-commands can be used.
%%%%%%

% The appendix (or appendices) are optional. For reviewing purposes
% additional 5~pages (double-column) are allowed (resulting in a maximum
% grand total of 10~pages plus one page containing only
% references). These additional 5~pages must be removed in the final
% version of an accepted paper.

\end{document}